\documentclass[twocolumn,pre,showpacs,showkeys,superscriptaddress]{revtex4-1}
\usepackage{graphicx}
\usepackage{soul}
\usepackage{color}
\usepackage{amsmath}
\usepackage{pgf}
\usepackage{amssymb}
\usepackage{float}
\usepackage{amsfonts}

\usepackage[greek,english]{babel}
\makeatletter
\newcommand*{\rom}[1]{\expandafter\@slowromancap\romannumeral #1@}

\begin{document}
\title{Multi-clustered chimeras in large semiconductor laser arrays with nonlocal interactions}

\author{J. Shena}
\affiliation{
Crete Center for Quantum Complexity and Nanotechnology, Department of Physics,
      University of Crete, P. O. Box 2208, 71003 Heraklion, Greece}
\affiliation{National University of Science and Technology MISiS, Leninsky prosp. 4, Moscow, 
      119049, Russia
}
\author{J. Hizanidis}
\email[corresponding author: ]{hizanidis@physics.uoc.gr}
\affiliation{
Crete Center for Quantum Complexity and Nanotechnology, Department of Physics,
      University of Crete, P. O. Box 2208, 71003 Heraklion, Greece}
\affiliation{National University of Science and Technology MISiS, Leninsky prosp. 4, Moscow, 
      119049, Russia
}
\author{P. H\"ovel}
\affiliation{Institut f{\"u}r Theoretische Physik, Technische Universit{\"a}t Berlin, Hardenbergstra\ss{}e 36, 10623 Berlin, Germany}
\affiliation{Bernstein Center for Computational Neuroscience Berlin, Humboldt-Universit{\"a}t zu Berlin,
Philippstra{\ss}e 13, 10115 Berlin, Germany}

\author{G. P. Tsironis}
\affiliation{
Crete Center for Quantum Complexity and Nanotechnology, Department of Physics,
      University of Crete, P. O. Box 2208, 71003 Heraklion, Greece}
\affiliation{National University of Science and Technology MISiS, Leninsky prosp. 4, Moscow, 
      119049, Russia
}
\affiliation{
Institute of Electronic Structure and Laser,
      Foundation for Research and Technology--Hellas, P. O. Box 1527, 71110 Heraklion,
      Greece}

\date{\today}

\begin{abstract} 

The dynamics of a large array of coupled semiconductor lasers is studied for a nonlocal coupling
scheme. Our focus is on chimera states, a self-organized spatio-temporal pattern of co-existing coherence and incoherence. 
In laser systems, such states have been previously found for global and nearest-neighbor coupling, mainly in small networks.
The technological advantage of large arrays has motivated us to study a system of 200 nonlocally coupled lasers
with respect to the emerging collective dynamics. The crucial parameters are the coupling strength,
the coupling phase and the range of the nonlocal interaction.  We find that chimera states with multiple
(in)coherent domains exist in a wide
region of the parameter space. We provide quantitative characterization for the obtained chimeras
and other spatio-temporal patterns.

\end{abstract}


\maketitle

\section{Introduction}
\label{sec:introduction}
Coupled lasers have been extensively studied in terms of nonlinear dynamics
\cite{25b,26b,27b,28b,29b} and synchronization phenomena \cite{30b,31b,32b}.
Most of these studies have been concerned with semiconductor laser arrays. They have been
demonstrated as sources that can produce high output power in a spatially coherent beam. 
Coupling between lasers may arise due to the overlap of the electric fields from each laser
waveguide or due to the presence of an external cavity \cite{Kozyreff,Fabian}.
In the latter case, a time delay is required for the mathematical modelling of the system. 
In general, most works on  laser arrays consider either global coupling, where 
each laser interacts with the whole system \cite{Rappel,Silber}, 
or local coupling, where each laser interacts with its nearest neighbors \cite{nikolis,winful2}.
The main property of those systems is that although the emission
from the individual elements is often unstable and even chaotic \cite{winful},
the total light output from the semiconductor array can be stable.  

In recent years, semiconductor laser networks have been studied in terms of a peculiar form
of synchronization called \textit{chimera states}. Since the first discovery of chimeras 
for symmetrically coupled Kuramoto oscillators in 2002 \cite{6b}, this counter-intuitive symmetry breaking phenomenon
of partially coherent and partially incoherent behavior has received enormous attention (see Ref.~\cite{7b}
and references within). In laser systems, chimeras were first reported both theoretically and experimentally
in a virtual space-time representation of a single laser system subject to long delayed
feedback~\cite{33b,34b}. Small networks of globally delay-coupled 
lasers have also been studied and chimera states were found for both small and large delays~\cite{Fabian,ROE16}.
Moreover, ``turbulent'' chimeras were recently observed and classified in large arrays of nonidentical laser arrays with nearest-neighbor interactions~\cite{turbulence}. There, the crucial parameters were the coupling strength and the relative optical frequency detuning between the emitters of the array.

\begin{figure}[t!]
\includegraphics[width=\linewidth]{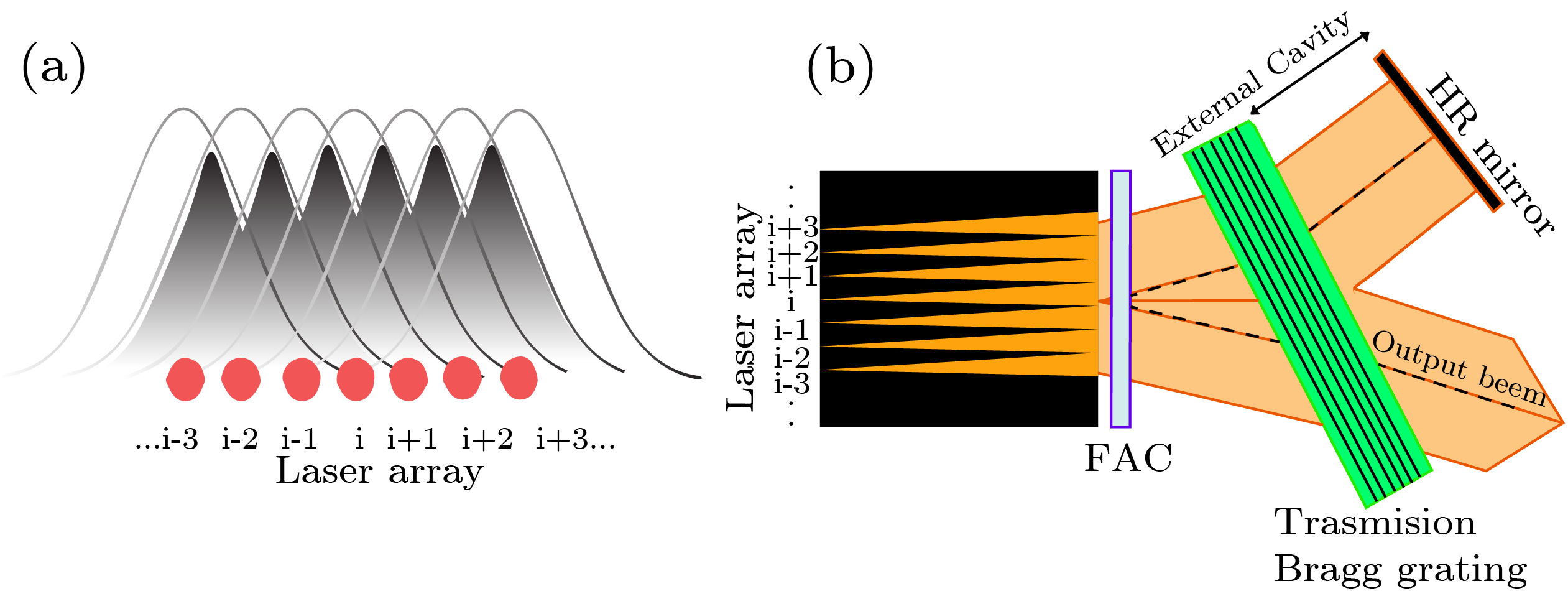}
\caption{Conceptual model of (a) the overlap of the electric fields in nonlocally coupled waveguide lasers, (b) a laser array coupled by a common highly-reflective (HR) mirror via an external cavity. (FAC stands for Fast-Axis Collimating lense). \label{fig_1}}
\end{figure} 

The experimental realization of laser arrays is challenging, but these devices have significant technological advantages: By achieving phase locking of the individual lasers we obtain a coherent and high-power optical source. In Ref.~\cite{devidson1} synchronization phenomena were studied in large networks with both homogeneous and heterogeneous coupling delay times. Moreover, in Ref.~\cite{devidson2} a new experimental approach to observe large-scale geometric frustration with 1500, both nonlocally and locally, coupled lasers was presented.
In the present work, we will focus on the intermediate case, i.e., nonlocal coupling. 
In laser networks this kind of coupling has never been attempted before, and with this paper we aim to fill this gap. In our study we use nonlocal coupling, where the crucial parameters for the observed dynamics are the strength, the phase and the range of the coupling. Our focus, in particular, is to identify the parameter regions where chimera states or other phenomena emerge and subsequently characterize them following a recently proposed classification scheme \cite{kevrekidis}.

\section{The Model}
\label{model}
In the present analysis, we consider a ring of 
$M=200$ semiconductors lasers of class B. Each node $j$ is 
symmetrically coupled with the same strength to its $R$ neighbors on either side (nonlocal coupling). 
The evolution of the slowly varying complex amplitudes ${\mathcal E}_{j}=E_j \exp{\left(i\phi_{j}\right)}$ (where $E_j$ is the amplitude and $\phi_{j}$ the phase of the electric field) and the corresponding population 
inversions $N_{j}$ are given by:

\begin{subequations} \label{eq1}
\begin{align}
\frac{d{\mathcal E}_{j}}{dt}&=(1+ia){\mathcal E}_{j}N_{j}+\frac{ke^{-i2C_{p}}}{2R}\sum_{l=j-  R}^{j+R}{\mathcal E}_{l} \\
\frac{dN_{j}}{dt}&=\frac{1}{T}\left(p-N_{j}-(1+2N_{j})|{\mathcal E}_{j}|^2\right), \quad j=1, \dots, M,
\end{align}
\end{subequations}
where all indexes has to be taken modulo $M$. $T$ is the ratio of the lifetime of the electrons in the excited level and that of the
photons in the laser cavity. Lasers are pumped electrically
with the excess pump rate $p=0.23$~\cite{Fabian}. The linewidth enhancement factor $a$ models the relation between the amplitude and the phase of the electrical field. We consider $a=2.5$, which is a typical value for semiconductor lasers. The coupling strength $k$ and the phase $C_{p}$ are the control parameters that are used to tune the collective dynamics of the system. This complex coupling coefficient 
models the important effect of a phase shift introduced as the electric field of one laser couples into another \cite{Katz}. Equations~\eqref{eq1} are a reduced form of the Lang-Kobayashi model in the limit where the delay of the external cavity tends to zero \cite{Fabian}. By adding shifting the coupling phase to $(C_p + \pi)$, we can obtain the model that describes the interaction of each field of semiconductor lasers in an array of waveguides  \cite{Rappel,Silber}.

\begin{figure*}
\includegraphics[width=0.9\textwidth]{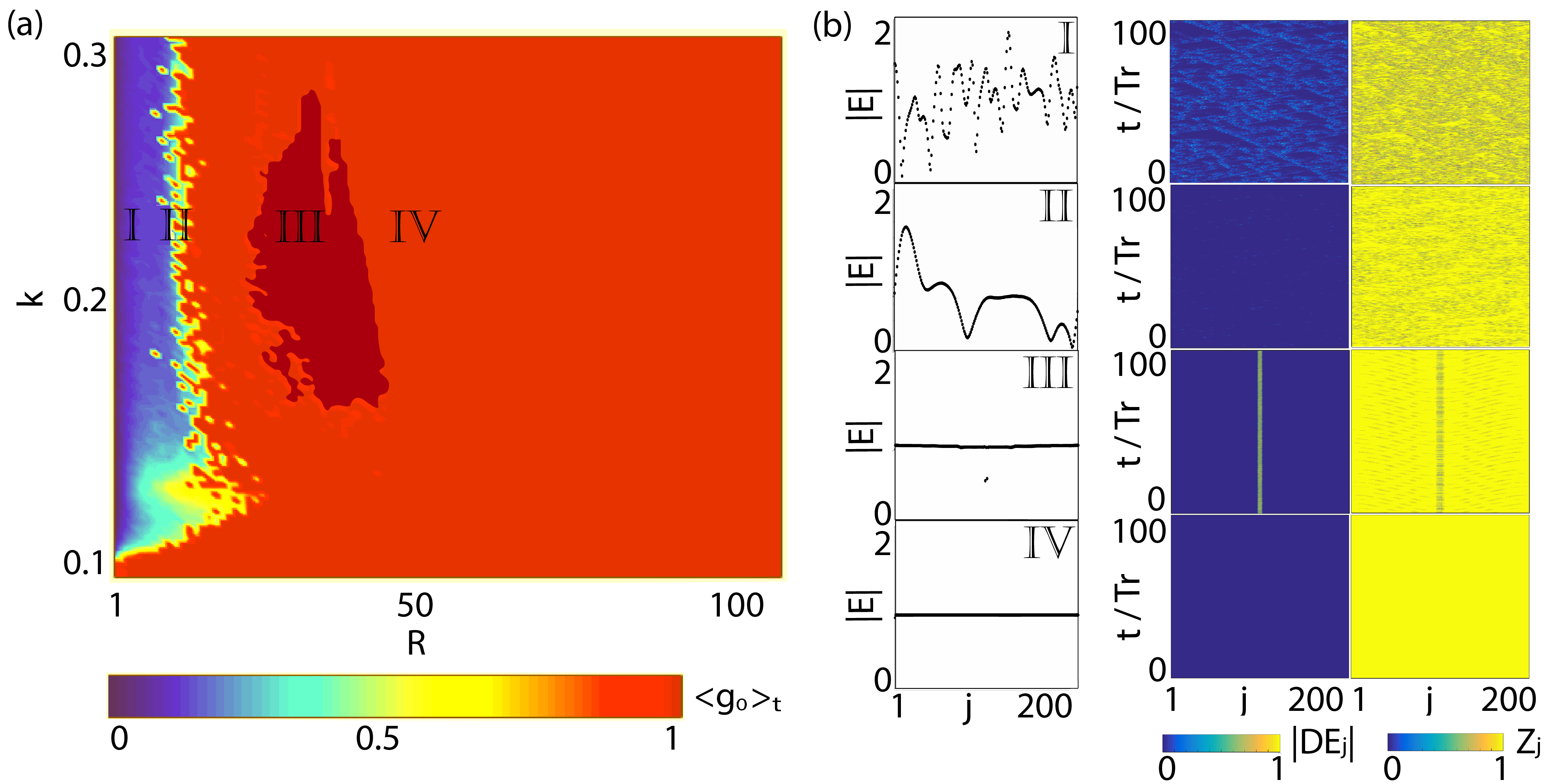}
\caption{(a): Dependence of the temporal mean $\left\langle g_{0}(t)\right\rangle$ on parameters $k$ and $R$. 
(b): Snapshot of the electric field (left), spatio-temporal evolution of the local curvature (middle), and the local order parameter (right) for fixed $k=0.21$ and four different coupling ranges: (\rom{1}) $R=2$, (\rom{2}) $R=9$, (\rom{3}) $R=29$, and
(\rom{4}) $R=50$. Other parameters: $T_{r}=392, p=0.23, a=2.5$, and $C_{p}=0$.\label{fig_2} }
\end{figure*} 

Physically, nonlocal coupling arises due to the overlap of the electric fields within a range 
of $R$ neighbor waveguides of lasers (see Fig.~\ref{fig_1}(a)). In this case, a portion of the electric field from one laser extends into the active region of its $2R$ neighboring lasers. The strength of this field extension decreases in space but for simplicity we assume a uniform coupling $k$ in every active region of $2R$ lasers.  
Another scheme corresponding to this type of coupling 
can be achieved by replacing all waveguides by a single external cavity where the length of it or the delay tends to zero (see Fig.~\ref{fig_1}(b)). In that case the converging lense coupler for all of lasers inside the cavity cannot 
converge all the $M$ beams of light in one beam and so a nonlocal coupling is a more realistic approach than an all-to-all coupling.    

For the initial conditions, the phases of the
individual lasers are randomly distributed along the complex
unit circle while amplitudes and inversions are chosen identical
for all lasers $ E_{j}(t=0)=\sqrt{p}$, $N_{j}(t=0)=0$. Moreover, the well known period $T_{r}=2\pi / \Omega$ of the individual 
laser relaxation oscillation frequency $\Omega=\sqrt{2p/T}$ will set the time scale of the system. 
In order to understand the effect all three control parameters, namely the coupling strength $k$, the coupling range $R$ and the coupling phase $C_{p}$, we split the problem into two parts: In the first part, the coupling phase is set to zero and the co-action of the coupling strength and range is studied. In the second part, the coupling phase is also considered and we will show that more complex phenomena like chimera states emerge. In the concluding section, we summarize our results and discuss open problems.

\section{Measures for phase and amplitude synchronization}
By using polar coordinates the characterization of the phase synchronization of our system can be done through the Kuramoto local order parameter~\cite{Omelchenko}:
\begin{equation} \label{eq4}
Z_{j}=\left |\frac{1}{2\zeta}\sum_{|l-j|\leq\zeta}e^{i\phi_{l}}\right |.
\end{equation}

We use a spatial average with a window size of $\zeta = 3$ elements.
A $Z_j$ value close to unity indicates that the $j$-th laser belongs
to the coherent regime, whereas $Z_j$ is closer to $0$ in the incoherent part.
This quantity can measure only the phase coherence and gives no information about the amplitude synchronization of the electric field.
For the latter, we will use the classification scheme presented in Ref.~\cite{kevrekidis} for spatial coherence,
which we have applied to other systems in recent works as well~\cite{turbulence,Ioannas}. 
In particular, we will calculate the so-called \emph{local curvature} at each time instance, by applying the absolute value of the 
discrete Laplacian $|DE|$ on the spatial
data of the amplitude of the electric field:
\begin{equation}\label{eq5}
|DE|_{j}(t)=\left|E_{j+1}(t)-2E_j(t)+E_{j-1}(t)\right|, \quad j=1,\dots, M. 
\end{equation}
In the synchronization regime the local curvature is close to zero while in the asynchronous
regime it is finite and fluctuating. Therefore, if $g$ is the normalized probability density function of $|DE|$, $g(|DE|=0)$ measures the relative size of spatially 
coherent regions in time. For a fully synchronized system $g(|DE|=0)=1$,
while for a totally incoherent system it holds that $g(|DE|=0)=0$. A value between $0$ and $1$
of $g(|DE|=0)$ indicates coexistence of synchronous and asynchronous lasers.

Note that the quantity $g$ is time dependent. Complementary to the local curvature we also calculate the spatial extent occupied by the coherent lasers which is given by the following integral:
\begin{equation}\label{eq6}
g_{0}(t)=\int_{0}^{\delta}g(t,|DE|)d|DE|,
\end{equation}
where $\delta=0.001$ is a threshold value distinguishing between coherence and incoherence, which is related to the system-dependent, maximum curvature.

\section{Collective dynamics}
In panel (a) of Fig.~\ref{fig_2}, the temporal mean of $g_{0}(t)$, averaged over $100T_{r}$, is plotted in the ($R,k$)-parameter space. There are four distinct regions: The blue area corresponds to the unsynchronized region, where $g_{0}(t)$ is close to zero
and is marked by the letter \rom{1}, and the red region, marked by the letter \rom{4}, refers to a stationary state where all lasers enter a fixed point and therefore $g_{0}(t)$ is close to unity,. 
Apart from those two well defined regions, there exist two more interesting ones for intermediate values of $g_{0}(t)$. The first one lies on the border between the incoherent and the stationary state and is marked by the letter \rom{2}, while the second region
exists within the stationary area and is marked by the letter \rom{3}.
Figure~\ref{fig_2}(b) shows the corresponding snapshot representations of the electric field (left), the spatio-temporal evolution of the local curvature (middle), and that of the local order parameter (right). 
Note that the local curvature has been normalized to its maximum value~\cite{kevrekidis}.

\begin{figure}
\includegraphics[width=\linewidth]{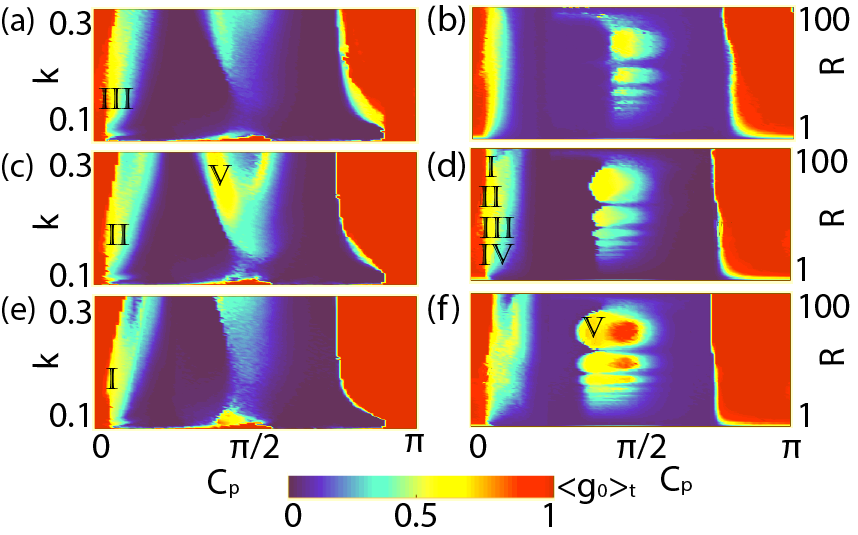}
\caption {Dependence of the temporal mean $\left\langle g_{0}(t)\right\rangle$ on parameters $k$ and $C_{p}$ for different values of nonlocal coupling range: (a) $R=40$, (c) $R=64$, and (e) $R=88$. Dependence on parameters $R$ and $C_{p}$ of the temporal mean $\left\langle g_{0}(t)\right\rangle$ for different values of the coupling strength: (b) $k=0.075$, (d) $k=0.15$, and (f) $k=0.225$. Other parameters: $T_{r}=392, p=0.23$, and $a=2.5$. \label{fig_4}}
\end{figure}

\begin{figure*}
\includegraphics[width=\textwidth]{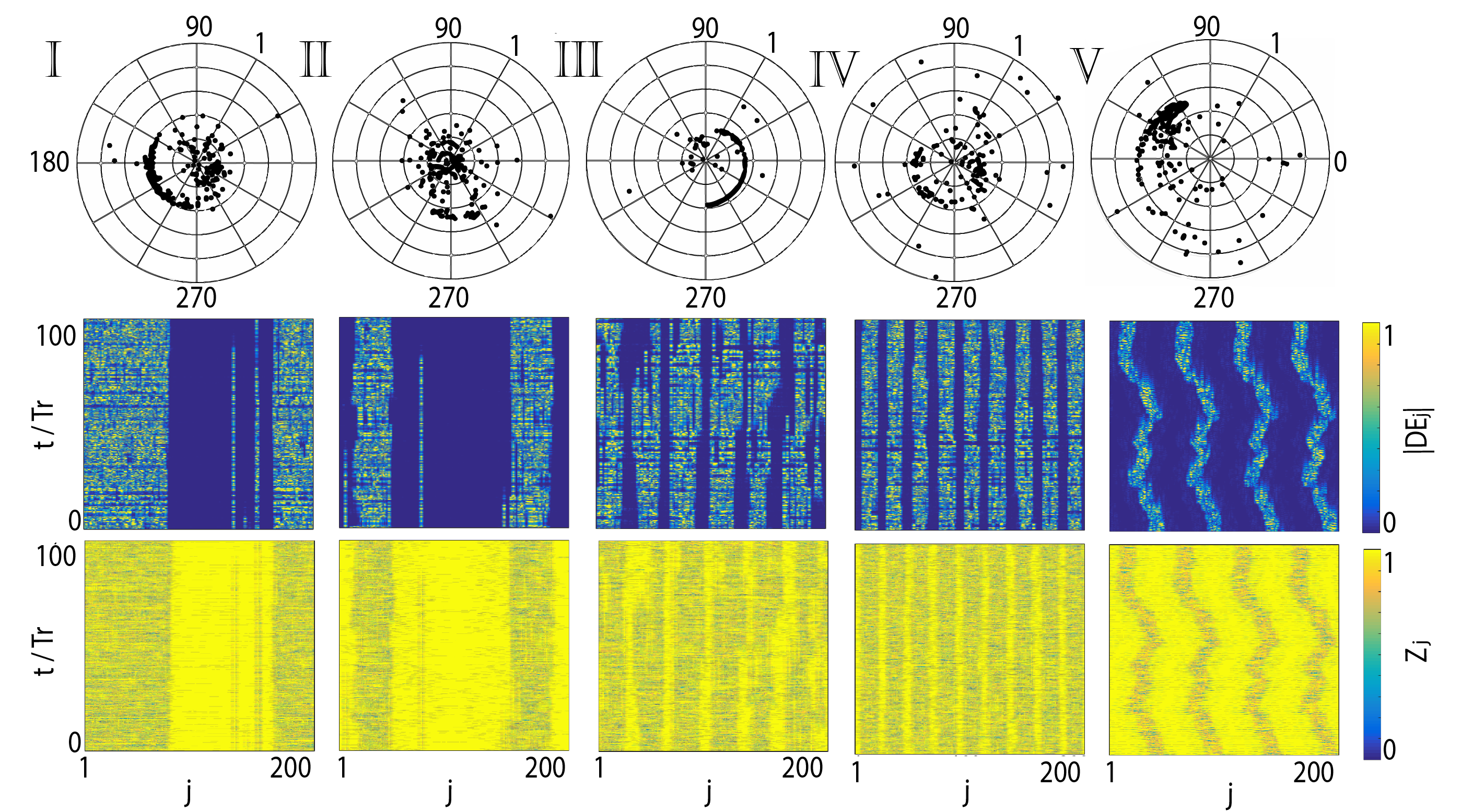}
\caption{The electric field in the complex unit circle (top), spatio-temporal evolution of the local curvature (middle), and spatio-temporal evolution of the local order parameter (bottom) for different coupling ranges and phases: (\rom{1}) $R=88, C_{p}=0.1\pi, k=0.15$ (\rom{2}) $R=64, C_{p}=0.06\pi, k=0.15$, (\rom{3}) $R=40, C_{p}=0.1\pi, k=0.15$, (\rom{4}) $R=27, C_{p}=0.1\pi, k=0.15$, and (\rom{5}) $R=64, C_{p}=0.4\pi, k=0.225$. Other parameters: $T_{r}=392, p=0.23$, and $a=2.5$. \label{fig_5} }
\end{figure*}

Moving from point \rom{1} to \rom{4}, the system goes from the incoherent state to the stationary one
through a wave-like spatial structure (point \rom{2}) and an almost fully stationary state (point \rom{3}).
In the incoherent state the lasers are desynchronized both in amplitude and in phase, which is depicted
in the local curvature and the local order parameter. With increase of the coupling range $R$, 
the temporal oscillations of the lasers tend to become closer in amplitude.  
This is reflected in the smooth wave-like structure of the electric fields and the discrete Laplacian which holds a
value close to zero. The corresponding phase oscillations are less coherent and this is evident by 
the blue areas in the order parameter spatio-temporal plot. 

Before entering the fully stationary state (\rom{4}) the systems undergoes another interesting
region where $g_{0}(t)$ is close, but less than one because of a deviation from the stationary state of two lasers (left panel of  \rom{3}), which holds for both the amplitude (middle) and the phase (right). In coupled systems, the phenomenon where one or more oscillators exhibit large amplitude oscillations whereas the rest are stationary, is called localized breather and has been intensively investigated in the past~\cite{Mackay,Chen}. 

For finite coupling phase $C_{p}$, the situation is much more complicated. 
By plotting the temporal mean of $g_{0}(t)$ in the $(C_{p},k)$-plane (Fig.~\ref{fig_4}(a), (c), and (e)) as well as in the $(C_{p},R)$-plane (Fig.~\ref{fig_4}(b), (d), and (f)) for various values of the coupling strength $k$ and the coupling range $R$, we can identify the existence of many patterns, among which chimera states, which we have marked with roman letters. 
Each chimera state is characterized by its multiplicity, i.e., the number of the (in)coherent regions also known as number of chimera clusters. Single chimeras (\rom{1}), as well as chimeras with two (\rom{2}), six (\rom{3}) and nine heads (\rom{4}) are observed. Moreover, localized oscillations and waves similar to those of Fig.~\ref{fig_2} are also found (not shown). Finally, ``turbulent'' chimeras~\cite{Shena} where the position of the (in)coherent regions changes in time and $g_{0}(t)$ oscillates irregularly complete the picture of the observed patterns (\rom{5}).   

More specifically, for nonlocal range coupling $R>10$ and coupling strength $k>0.05$,  we can distinguish different regions in terms of the coupling phase value. Below those two values the interaction is so weak that each laser behaves like an uncoupled one (see Fig.~\ref{fig_4}, lower left corners of all panels). Around the region $C_{p}\approx 0$ and the region $C_{p}\approx \pi$ the case of full synchronization is most prominent, where $\mathcal E_{i} =\mathcal  E_{j}$ holds for all lasers. The opposite situation of full asynchrony where both amplitude and phase exhibit incoherent behaviour appears around the regions $C_{p}\approx \pi/4$ and $C_{p}\approx 3\pi/4$. On the boundary between full synchronization and asynchrony lies a small area where the chimeras arise.

 Figure~\ref{fig_5} shows typical snapshots of multi-clustered
 chimera states of the electric field in the complex unit circle (top panel), the spatio-temporal evolution of the local curvature (middle panel) and the spatio-temporal evolution of the local order parameter 
(bottom panel) for points \rom{1}-\rom{4}. We observe that the decrease of $R$ yields additional chimera heads both in amplitude and in phase.
Finally, around the region where $C_{p}\approx \pi/2$ turbulent chimeras appear (Fig.~\ref{fig_5}, \rom{5}).

\section{Conclusions}\label{real}
In conclusion, multi-clustered chimera states have been obtained and characterized in large arrays
of semiconductor class B lasers with nonlocal interactions.
The observed chimeras display the coherence and incoherence patterns in both the amplitude and phase
of the electric field and can be both stationary or ``turbulent'', where the size and position of the 
(in)coherent clusters vary in time.
In addition, other spatiotemporal dynamics including wave-like spatial structures and spatially localized oscillations (breathers)
are possible.
The crucial parameters for the collective behavior are the complex coupling strength and the nonlocal coupling range. 
By applying recently presented measures for spatial coherence we have identified and classified
the emerging dynamics in the relevant parameter spaces.
Our study addresses the effect of nonlocal coupling in large laser arrays for the first time, 
providing a direction for various technological applications.
For future studies it would be worthwhile to explore the influence of noise and anisotropy in the laser pump 
power.

\section{Acknowledgments}
This work was partially supported by
 the Ministry of Education and Science of the Russian Federation in the framework of the Increase Competitiveness Program of NUST ``MISiS'' (No. K2-2015-007) and the European Commission under project NHQWAVE (MSCA-RISE 691209). PH acknowledges support by Deutsche Forschungsgemeinschaft in the framework of SFB 910. 
 Moreover, the authors would like to thank V. Kovanis for fruitful discussions.

\end{document}